# PC GUIDED AUTOMATIC VEHICLE SYSTEM

M.A.A. Mashud [#1], M. R. Hossain [#2], Mustari Zaman [#2] and M.A. Razzaque [#1]

[#1] Dept. of Applied Physics, Electronics & Communication Engineering, Islamic University, Kushtia-7003, Bangladesh.
[#2] Dept. of Information & Communication Engineering, Rajshahi University, Rajshahi, Bangladesh.

## ABSTRACT

*The main objective of this paper is to design and develop an automatic vehicle, fully controlled by a computer system. The vehicle designed in the present work can move in a pre-determined path and work automatically without the need of any human operator and it also controlled by human operator. Such a vehicle is capable of performing wide variety of difficult tasks in space research, domestic, scientific and industrial fields. For this purpose, an IBM compatible PC with Pentium microprocessor has been used which performed the function of the system controller. Its parallel printer port has been used as data communication port to interface the vehicle. A suitable software program has been developed for the system controller to send commands to the vehicle.*

## 1. INTRODUCTION

This paper presents a computer controlled automatic vehicle system. This vehicle can move in a pre-determined path without any intervention from a human operator. The developed system has two basic units: (1) a control unit consisting of a personal computer (PC) containing the parallel printer port and a suitable software program which enables the vehicle to operate automatically and (2) a vehicle control unit.

To establish communication between the control unit and the vehicle unit, there are two modes of data transfer: parallel and serial. Parallel communication uses multiple wires to represent the information. Within a microcomputer, data are transferred in parallel mode, because it facilitates the fastest data transfer. But for transferring data over a long distance, parallel data transmission requires too many wires. On the other hand, in serial communication system a series of pulses are transmitted on a single wire. Hence for long distance data communication, the serial data transfer is preferred, as it is simple to control and also cost-effective.

A personal computer is available at a cheaper price now-a-days and it is easier to interface external devices with it. In this work, an IBM PC with Pentium microprocessor is used that performs the function of a system controller [1].

There are several ways of inputting and outputting data i.e. as I/O interfacing. The PC contain parallel printer port, serial port, keyboard port, mouse port, USB port, 8255 port, game port etc. among these the parallel printer port is the most common way of interfacing a normal PC to the outside world and serial port is one of the oldest used for interfacing. Parallel printer port has a distinct advantage to the others. By changing the parameters of its control register, it can be used for bi-directional data transmission in all the modes of data transfer: SPP, EPP and ECP. In the case of serial port this is not possible. Moreover the data transmission rate of serial port is fixed at RS-232 standard and cannot be modified. For this reasons, in the present work, the parallel printer port has been used to interface the vehicle unit with the control unit.

Simple, locally available and low cost components have been used in the present work to develop the motor driving circuit of the vehicle. Simple motor driving circuits have been used in the vehicle to move it in the forward or backward direction noiselessly and efficiently by using a unipolar power supply. Stepper motor has been attached to the front wheels to turn the vehicle in the desired direction (left or right).

## 2. SYSTEM DESIGN

The implemented hardware provides necessary interface between the microcomputer and the system to be controlled. The developed system is fully software controlled. The speed of communication and switching are determined by the hardware. The complete system that comprises of the control unit and the remote vehicle unit is divided into the following four parts:

1. Microcomputer and its parallel printer port.
2. Software for data transmission from the control unit through parallel printer port.
3. Motor driving unit of the vehicle.
4. Mechanical parts of the vehicle.

For the purpose of transmitting and receiving data the parallel printer port is used. The receiver and other timing circuits were arranged according to required standard. In order to make the transmission rate independent and also to control the rate through external timing devices, the parallel printer is used. The developed system (vehicle unit) is interfaced with the control unit (PC) through printer port. The simplified block diagram of the proposed system is depicted in Figure 1.

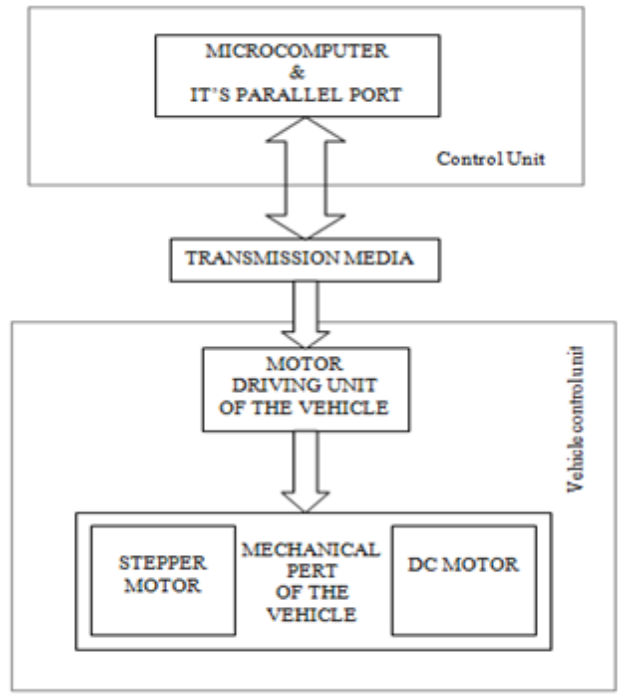

Figure 1. A simple block diagram of the proposed system

## 2.1. Motor Driving Circuit

The main mechanical parts of the vehicle is steering control circuit. For steering control we have used stepper motor [2]. The complete circuit diagram for steering control, connected with parallel printer port and stepper motor is shown in Figure 2.

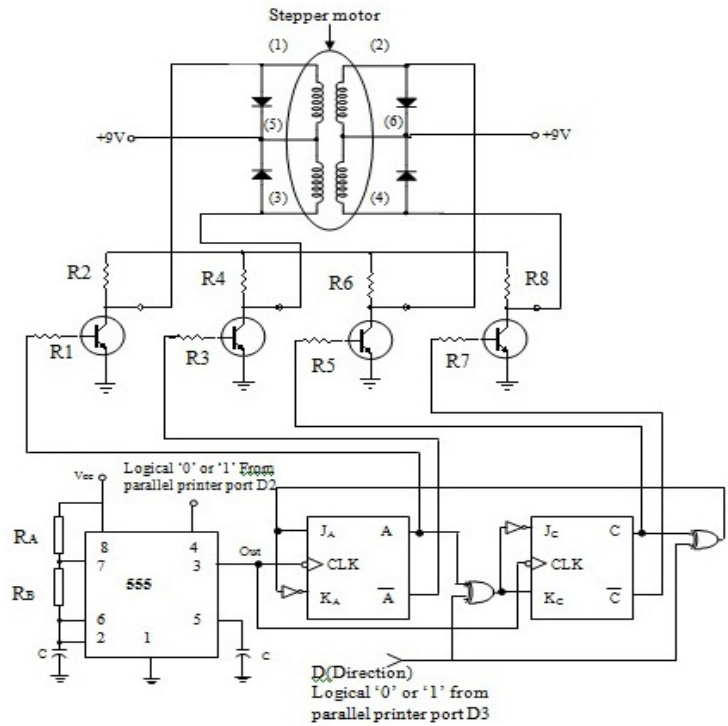

Figure 2. Steering control circuit of the vehicle

When the signal at pin-4 of NE555 IC is high (logic '1'), then the astable multivibrator circuit is active and it gives continuous clock pulse to the synchronous counter clock input terminal. When the direction input is HIGH (logic '1') the counter count pulses in clockwise direction. Therefore the stepper motor moves in clockwise rotation. But when the direction input is LOW (logic '0') then the counter count pulses in counter clockwise direction. The stepping motor sequence is shown in table 1.

Table 1. Stepping motor sequence.

| Direction bit | Counter output | | | | Counting sequence (direction of motor rotation) |
|---|---|---|---|---|---|
| | A | $\overline{A}$=B | C | $\overline{C}$=D | |
| HIGH (Logic '1') | 1 | 0 | 1 | 0 | Clockwise |
| | 0 | 1 | 1 | 0 | |
| | 0 | 1 | 0 | 1 | |
| | 1 | 0 | 0 | 1 | |
| | 1 | 0 | 1 | 0 | |
| LOW (Logic '0') | 1 | 0 | 1 | 0 | Counter clockwise |
| | 1 | 0 | 0 | 1 | |
| | 0 | 1 | 0 | 1 | |
| | 0 | 1 | 1 | 0 | |
| | 1 | 0 | 1 | 0 | |

## 2.2. Parallel Printer Port

The parallel port is most commonly used port for interfacing homemade projects [3]. This port consists with the input of up to 9 bits and the output is of 12 bits at any one given time. Thus it requires minimal external circuitry to implement many simpler tasks. It is located at the back of our PC as a D-Type 25 Pin female connector is shown in Figure 3. The original IBM-PC has Parallel Printer Port with a total of 12 digital outputs and 5 digital outputs accessed via 3 consecutive 8-bit ports in the processor's I/O space. 8 output pins accessed via the DATA Port. It has 5 input pins (one inverted) and 4 output pins (three inverted) that accessed via the STATUS Port and CONTROL Port respectively. The remaining 8 pins are grounded. It' may also be a D-Type 25 pin male connector. This is a serial port named RS-232 port and totally incompatible.

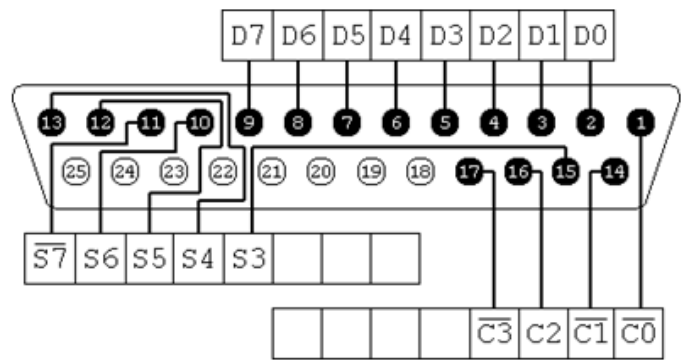

Figure 3. Parallel printer port pins configuration [4].

## 2.3. Interfacing Circuit

The interfacing circuit is design for an I/O device to determine primarily by the instructions to be used for data transfer [5]. Each I/O device is assigned a binary address, called a device address or port number, through its interfacing circuit. A peripheral-mapped I/O or a memory-mapped I/O is needed to interface an I/O device. While the microprocessor executes a data transfer instruction for an I/O device, it places the appropriate address on the address bus and sends the control signals according to the type of operation to be performed which enables the interfacing device and transfers data. The data bytes are transferring by of I/O instructions technique is called peripheral-mapped I/O. when the data bytes are transferred by using memory related data transfer instructions, the technique is called memory-mapped I/O. however the basic concepts in interfacing I/O devices are similar in both methods. In this project, peripheral-mapped I/O technique has been used. The block diagram of Figure 4 illustrate a simple peripheral-mapped I/O interfacing technique for an I/O device, where the device address bus and control signals are used as follows:

(i) The address bus is decoded to generate a unique pulse corresponding to the device address on the bus, which is called device address pulse.
(ii) The device address pulse is combined with the control signal for generating a device select (I/O select) pulse which is generated only when both signal are asserted.
(iii) Then the device select pulse is used to activate the interfacing device (I/O port).

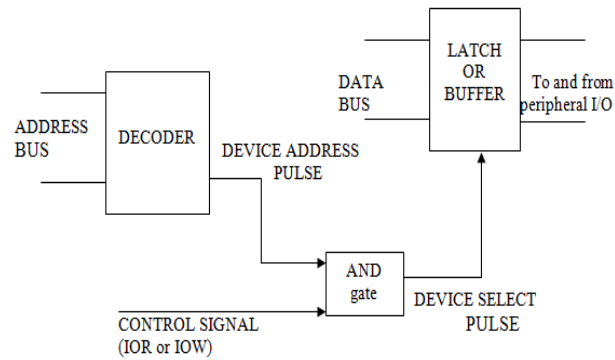

Figure 4. A simple block diagram of I/O interface

The basic input output device is a set of three-state buffers. The basic output device is consisting of data latches. The term IN presents for moving data from the I/O device into the microprocessor, and the term OUT presents for moving data out of the microprocessor to the I/O device.

The three-state buffers are used to construct an 8-bit input port. The external TTL logic data are connected to the input of the buffers. The output of the buffers connected to the data bus. While the microprocessor executes an IN instruction then the I/O port address is decoded for generating the logic 0 on SEL. Logic 0 placed on the output control inputs (1G and 2G) of the 74ALS244 buffer causes the data input connections (A) to be connected to the data output (Y) connections. If a logic-I is place on the output control inputs of the 74ALS244 buffer, the device enters the three-sate high impedance mode that effectively disconnect the switches from the data bus. The basic input circuit is essential for appearing any time that input data are interfaced to the microprocessor.

The output interfacing circuit receives data from the microprocessor and must usually hold it for some external device. The buffers found in the input device are often built into the I/O device. Figure 4 shows how eight bit data line connected to the microprocessor through a set of eight data latches are needed to hold the data because when the microprocessor executes an OUT instruction, the data are only present on the data bus for less than 0.1 microseconds.

## 2.4. Motors

Two types of motors have used in this work [6, 7]. For the preciously control of steering, the stepper motor is used. For movement in forward and backward noiselessly DC motor is used as shown in Figure 5.

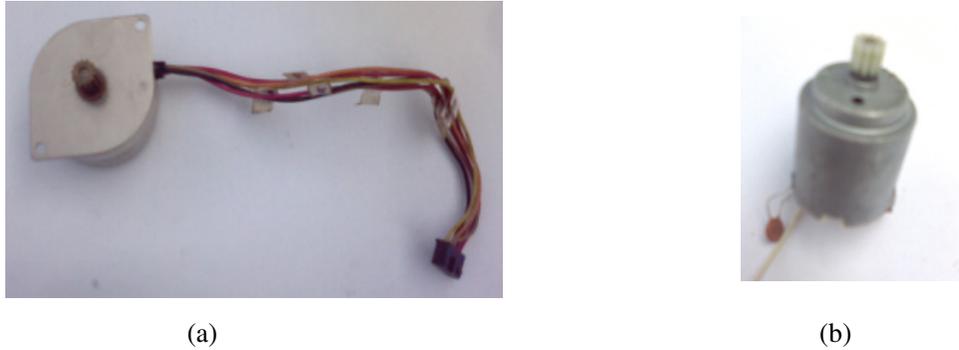

(a)  (b)

Figure 5. (a) Stepper Motor and (b) DC motor

## 3. SOFTWARE

The developed system consists by three parts. Among the three one is the personal computer. Computer needs clear-cut instructions to tell them what to do, how to do, and when to do. The microprocessor needs a set of instructions to perform a task is called a program, and a group of meaningful and valid program is called software. Software activities are those associated with the successful development to operate of the computing function and processing anything that drives the computer. Of course in the development of software much hardware equipment is not involved [8, 9].

*Algorithm of the Software:*

Since the software dominants the system function, hence it has been developed for proper operation of the system. The software sends command to the vehicle for reaching the destination. The algorithm of the developed software is given bellow:

1. Read a command from the keyboard
2. If scan = UP ARROW
        Move forward
    Else if scan = RIGHT ARROW
        Move right
    Else if scan = LEFT ARROW
        Move left
    Else if scan = BACK ARROW
        Move backward
    Else if scan = alphabetic 'S'
        Stop
3. If scan = END
        Return
    Else go to step two

## 4. RESULTS AND DISCUSSION

In the present work, an attempt has been made to design and develop an automatic vehicle system, capable of moving in a predetermined path without any intervention from a human operator. Here, the computer (PC) works as the controller of the remote vehicle, which is interfaced with it through the parallel printer port. For the system the hardware portion was first design and implemented and then software was developed to control the whole system. The software permits the computer (PC) to transmit the commands to the remote vehicle in serial binary form through the parallel printer port to the vehicle.

An IBM parallel printer port is used to interface the developed automatic vehicle system. The printer port has been used as a bi-directional data communication port in the SPP mode whose base address is assigned as 378h. The port addresses of parallel printer port are fixed to 378h to 37Fh for LPT1. For the joint venture between Intel, Xircom & Zenith Data Systems enhanced parallel port (EPP) was designed. Firstly specified the EPP Ports in EPP1.7 standard, and then included in the IEEE 1284 Standard released in 1994. Two Standards of EPP are EPP1.7 and EPP1.9. The typical transfer rate of EPP is the order of 500KB/S to 2MB/S. In order to perform a value exchange of data using EPP handshake must be followed. Since the hardware does all the work, this handshake only requires to be used for the hardware and not for software as the case with SPP. For initiating an EPP cycle, the software requires to perform only one I/O operation to the relevant EPP register. EPP data write cycle may be explained on the following way:

1. Program writes to EPP Data Register. (Base +4)
2. nWrite is placed low. (Low indicates write operation)
3. Data is placed on Data Lines 0-7.
4. nData Strobe is associated if Wait is Low. (O.K. to start cycle)
5. Host waits while Acknowledgment for nWait going high (O.K. to end cycle)
6. De-asserted the nData Strobe.
7. EPP Data Write Cycle Ends

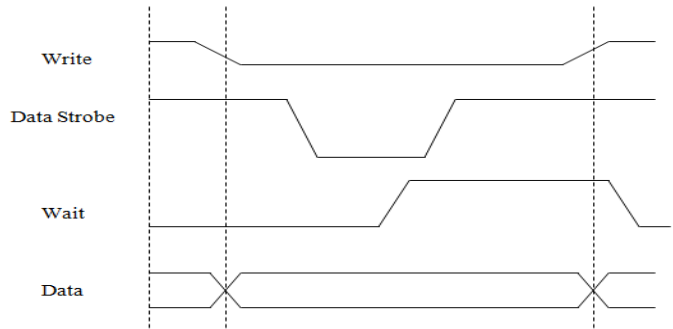

Figure 6. Enhanced Parallel Port Data Write Cycle

The proposed system was tested successfully. The system was tested in hazards place and found good performance. The general specifications of the designed system are depicted in Table 2.

Table 2: General specifications of the designed system

| | |
|---|---|
| Average velocity | 14 cm / sec |
| Maximum Rotational angle | 45º |
| Minimum Rotational angle | 1.8º |
| Transmission Media | Wire |
| Length | 35.5 cm |
| Width | 14.1 cm |
| Weight | 1.00 K.g |

International Journal on Cybernetics & Informatics (IJCI) Vol. 3, No. 6, December 2014

## 5. CONCLUSIONS

The proposed system is capable of moving in a predetermined path without any intervention from a human operator. Here, the computer (PC) works as the controller of the remote vehicle, which is interfaced with it through the parallel printer port. For the system the hardware portion was first design and implemented and then software was developed to control the whole system. The software permits the computer (PC) to transmit the commands to the remote vehicle in serial binary form through the parallel printer port to the vehicle. In this work hardware designs are simple, reliable, and are developed by using low cost locally available components. The motor driving circuits in the present work reverse the motor direction noiselessly and efficiently with the use of a unipolar power supply. The developed system in the present work is fully software controlled. The use of PC offers a simple, flexible and reliable solution of the system. It seems that the proposed system will provides a well structured controlled organization.

**Authors**

**Md. Abdullah-Al-Mashud** was born on Nov.15, 1980 in kushtia, Bangladesh. He received the B.Sc (Hons) degree and M.Sc degree in Applied Physics, Electronics and Communication Engineering (APECE) from Islamic University, Kushtia, Bangladesh in 2003 and 2004 respectively. He works as a faculty member in the department of APECE, Islamic University, Bangladesh. His current interest is microprocessor / microcontroller 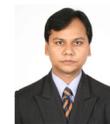 applications in control, automation, medical instruments, environmental monitoring, low cost electronic systems, Medical Image Processing. His work has produced 14 peer-reviewed scientific International and National Journal papers. He has published 07 papers in National and International Conferences.

**Md. Reaz Hossain** was born on Jan 17, 1976 in Dhaka, Bangladesh. He received the M.Sc degree in Applied Physics & Electronics from Rajshahi University, Bangladesh. He is an Assistant Professor in the Department of Information and Communication Engineering, Rajshahi University, Bangladesh. He has produced 11 peer-reviewed scientific International 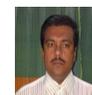 and National Journal papers. His current interest is Bioinformatics, Robotics, Network Security System, automation, medical instruments.

**Mustari Zaman** was born in Rajshahi, Bangladesh. She received the M.Sc degree in Applied Physics & Electronics from Rajshahi University, Bangladesh. She is a Assistant Professor in the Department of Information Science and Communication Engineering, Rajshahi University, Bangladesh. Now she is a Ph.D fellow at the Department of Electrical and Computer Engineering at Curtin University, Australia. She has produced 13 peer-reviewed scientific International and National Journal papers. He has published 02 abstracts in National and International Conferences. 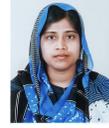

**Md. Abdur Razzaque** was born in May. 10, 1976 in Meherpur, Bangladesh. He received the M.Sc degree in Applied Physics, Electronics and Communication Engineering (APECE) from Islamic University, Kushtia, Bangladesh in 2000. He works as a Assistant Professor in the department of APECE, Islamic University, Bangladesh. His current interest is signal processing, Intelligent Controls and Mobile telecommunications. His work has produced 03 scientific l Journal papers. 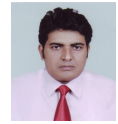